\documentclass[twocolumn,aps,prl,showpacs,letterpaper,superscriptaddress]{revtex4}
\usepackage[dvips]{graphicx,color}
\begin{document}

\title{Kondo effect in single atom contacts: the importance of the atomic geometry}

\author{Lucia Vitali}
\affiliation{Max-Planck-Institut f\"ur Festk\"orperforschung,
             Heisenbergstr.1, D-70569 Stuttgart, Germany}

\author{Robin Ohmann}
\affiliation{Max-Planck-Institut f\"ur Festk\"orperforschung,
             Heisenbergstr.1, D-70569 Stuttgart, Germany}

\author{Sebastian Stepanow}
\affiliation{Max-Planck-Institut f\"ur Festk\"orperforschung,
             Heisenbergstr.1, D-70569 Stuttgart, Germany}
\affiliation{Centre d'Investigacions en Nanociència i
Nanotecnologia (CIN2-ICN), E-08193 Bellaterra, Barcelona, Spain}

\author{Pietro Gambardella}
\affiliation{Institució Catalana de Recerca i Estudis Avançats
(ICREA)E-08193 Barcelona, Spain} \affiliation{Centre
d'Investigacions en Nanociència i Nanotecnologia (CIN2-ICN),
E-08193 Bellaterra, Barcelona, Spain}

\author{Kun Tao}
\affiliation{Max-Planck-Institut f\"ur Mikrostrukturphysik,
Weinberg 2, D-06120 Halle, Germany}

\author{Renzhong Huang}
\affiliation{Max-Planck-Institut f\"ur Mikrostrukturphysik,
Weinberg 2, D-06120 Halle, Germany}

\author{Valeri S. Stepanyuk}
\affiliation{Max-Planck-Institut f\"ur Mikrostrukturphysik,
Weinberg 2, D-06120 Halle, Germany}

\author{Patrick Bruno}
\affiliation{Max-Planck-Institut f\"ur Mikrostrukturphysik,
Weinberg 2, D-06120 Halle, Germany}
\affiliation{European
Synchrotron Radiation Facility BP 220, F-38043 Grenoble Cedex,
France}

\author{Klaus Kern}
\affiliation{Max-Planck-Institut f\"ur Festk\"orperforschung,
             Heisenbergstr.1, D-70569 Stuttgart, Germany}
\affiliation{Institut de Physique des Nanostructures, Ecole
Polytechnique F\'ed\'erale de Lausanne (EPFL), CH-1015 Lausanne,
Switzerland}

\date{\today}

\begin{center}
  \begin{abstract}
Co single atom junctions on copper surfaces are studied by
scanning tunneling microscopy and \emph{ab-initio} calculations.
The Kondo temperature of single cobalt atoms on the Cu(111)
surface has been measured at various tip-sample distances ranging
from tunneling to the point contact regime. The experiments show a
constant Kondo temperature for a whole range of tip-substrate
distances consistently with the predicted energy position of the
spin-polarized d-levels of Co. This is in striking difference to
experiments on Co/Cu(100) junctions, where a substantial increase
of the Kondo temperature has been found. Our calculations reveal
that the different behavior of the Co adatoms on the two Cu
surfaces originates from the interplay between the structural
relaxations and the electronic properties in the near-contact
regime.

   \end{abstract}
\end{center}

\maketitle

The electron transport properties in a circuit, whose dimensions
are reduced to single atom/molecule contact, is dominated by the
quantum character of matter \cite{cuniberti05} and requires a deep
understanding of its nanometer scale properties. Additionally, the
electron transport through such a junction is strongly influenced
by the coupling of the orbitals of the electrodes to the bridging
molecule or atom \cite{tao08}. Accessing the correct information
of the geometrical arrangement of the contact is fundamental to
the interpretation of the experimental data. This is often
challenging, despite the continuous progress in understanding the
electron transport through nanometer scale junctions
\cite{scheer98, yanson98, pascual95, lortscher07, burgi99, neel07,
termirov08, xiao04}.
 Among other techniques such as mechanical
controlled break-junction \cite{scheer98, yanson98, lortscher07}
or electro-migration \cite{park99}, a junction achieved with the
tip of a scanning tunneling microscope (STM) on a metal surface
\cite{gimzewski95, burgi99,neel07, termirov08} has been proven as
a valuable tool to target and select the substrate configuration
before and after the contact is formed. Nonetheless, a general
picture on the influence of atomistic order on the electron
conductance at nanometer scale junctions is, at present, still
missing.

Co adatoms on copper surfaces constitute an ideal system for
studying the electron transport through nanometer scale junctions.
The interplay of the unpaired electrons of single Co atoms and the
free electron states on copper surfaces leads to the formation of
a narrow electronic resonance at the Fermi level known as the
Kondo resonance. This has been extensively characterized in
tunneling conditions \cite{madhavan98, wahl04, knorr02}. One main
result of these studies is the evidence that on different
supporting surfaces the width of this resonance coincides with a
change of the occupation of the d electron levels of the Co adatom
\cite{wahl04}. This inherently reflects a varied coupling of the
magnetic impurity to the metal substrate. Indeed, a simple model
has been suggested to relate the energy position of the electron
d-levels of the impurity and the atomic arrangement in close
proximity of the Co adatom \cite{wahl04, knorr02}. Specifically, a
narrower Kondo resonance is observed for Co adatoms on Cu(111)
than on the Cu(100) surface in agreement with a shift in energy of
the d-levels i.e with an increase in their occupation from the
first to the second surface \cite{knorr02, wahl04}. Due to this
dependence, the width of the Kondo resonance is then a good
reference parameter to characterize the influence of the tip at
reduced distances. However, it is a priori not evident if the
width of the Kondo resonance will follow a trend similar to the
one observed in tunneling configuration on various surfaces also
when the tip is approached to the point contact configuration.

\par\

In order to address this question we measured the current-voltage
characteristics at different tip-sample distances ranging from
tunneling to point contact on individual Co adatoms on a Cu(111)
surface. As will be shown in the following, the width of the Kondo
resonance is practically constant on this surface at all
tip-substrate distances in an apparent contradiction with the
previously reported results on Cu(100) \cite{neel07}.

Based on \emph{ab-initio} theoretical calculations aimed to
determine the electronic and magnetic properties of these two
systems, we will show that the opposing results observed on the
two copper surfaces are not contradicting but demonstrate nicely
the determining influence of the local atomic structure on the
transport properties of a nanoscale junction.

\par\

The experiments were performed  using a home built scanning
tunneling microscope operated at 6K in ultra high vacuum (UHV)
with a base pressure of 1*10$^{-11}$mbar. The Cu(111) single
crystal has been cleaned in UHV by cycles of Ar ion sputtering and
annealing. Co single atoms have been deposited from a thoroughly
degassed Co wire wound around a tungsten filament on the Cu
surface at ~20K. This resulted in a coverage of about 10$^{-3}$ML
of isolated immobile Co adatoms.  The STM tip, chemically etched
from tungsten wire, was treated in vacuo by electron field
emission and soft indentation into the copper surface. This
assured a spectroscopically featureless tip near the Fermi energy.
Given this preparation, the tip was most likely covered by copper
atoms deriving from the substrate.

\par\

The inset in figure \ref{fig1} shows the conductance of a single
Co adatom at various tip-substrate displacements. This has been
achieved by recording the current while approaching the tip
towards the atom, in open feedback loop  conditions. As the tip
substrate distance is reduced the current increases smoothly from
the tunneling to the point contact regime following the
exponential dependence with the tip-substrate distance (z)
characteristic of the electron tunneling process
I(z)=I$_0$exp(-Az) (where A is proportional to the work function
of tip and substrate). Fitting the experimental curve, we obtained
a work function for the Co/Cu(111) system of 5 eV. As the point
contact regime is reached the current is found to exhibit a
characteristic quantization plateau with only a weak dependence on
the distance. The plateau is observed to be 1G$_0$ where G$_0$ is
the conductance quantum G$_0$=2e$^2$/h (h is Planck constant) in
agreement with studies on Co/Cu(100)\cite{neel07}. Topographic
images acquired before and after the tip was approached and
retracted from the point contact configuration confirm that the
contact region as well as the tip have not changed during the tip
displacements.

\begin{figure}[t]
 \centering
   \includegraphics[keepaspectratio=true,width=0.9\linewidth]{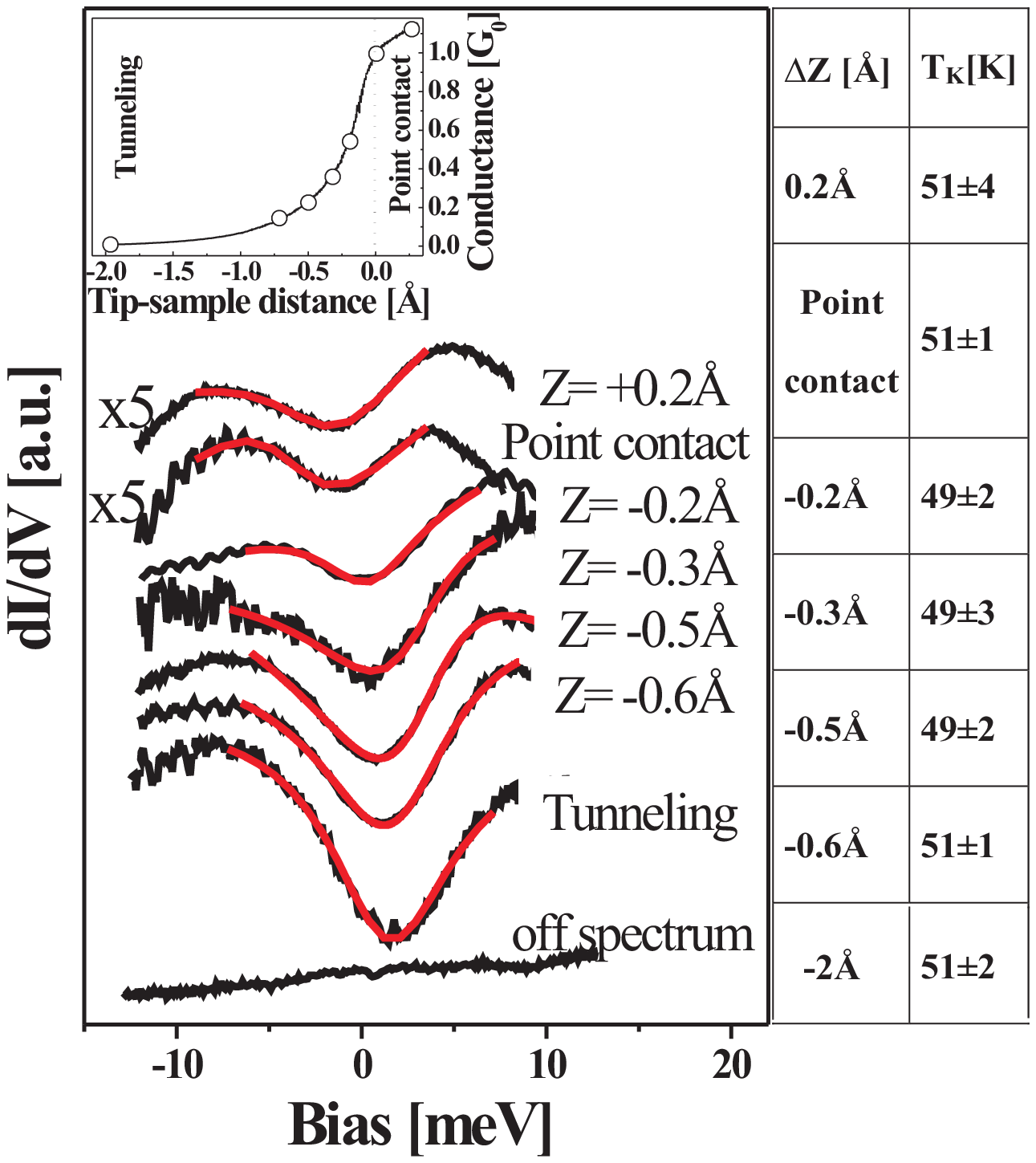}
   %bb=186 261 518 691
   \caption{\label{fig1} (color on line) Conductance and dI/dV spectra
   for isolated Co atoms on a Cu(111) surface achieved
   at different tip-substrate distances $\Delta$Z from tunneling to point
   contact. In inset a representative current vs tip-displacement is shown.
    dI/dV spectra have
been recorded at the position indicated by a circle in the inset.
The curves are normalized to the tunneling current at the tip
height location and vertical offset has been added for a better
visualization. The Kondo temperature T$_K$ given on the right side
of the image has been obtained fitting the curves with a Fano line
shape (red line). }

 \label{fig1}
\end{figure}

\par\
Information on the Kondo resonance have been obtained recording
the dI/dV spectra  on top of the Co adatom at various
tip-substrate displacements. In figure \ref{fig1}, we report the
spectra obtained at the tip-substrate separation indicated by the
circles in the inset. The current versus voltage is measured with
a lock-in amplifier applying a voltage modulation in the range of
1 to 0.1mV (rms). All the curves obtained in the range from the
initial tunneling ($\Delta$Z=-2\AA) to the point contact
($\Delta$Z=0.2\AA) condition show a characteristic dip in the
local density of states at an energy close to the Fermi level.
This dip, which is due to the Kondo resonance can be characterized
according to its width $\Delta$E, which is proportional to the
Kondo temperature T$_K$, $\Delta$E=2k$_B$T$_K$, where k$_B$ is the
Boltzmann constant \cite{madhavan98, wahl04}. The Kondo
Temperature can be extracted from these curves by fitting the
experimental spectra with a Fano line function according to
dI/dV$\propto$(q+$\epsilon$)$^2$/(1+$\epsilon^2$), with
$\epsilon$=(eV-$\epsilon$$_K$)/k$_B$T$_K$ where  q and
$\epsilon_K$ define the asymmetry of the curve and the energy
position of the resonance with respect to the Fermi energy
\cite{fano61}. The fitted Kondo temperature T$_K$ is reported in
figure \ref{fig1} for each sampled tip position. As can be seen,
the Kondo temperature for the Co on Cu(111) system is constant,
within the experimental error, from the tunneling to the point
contact regime.

The observed behavior of the Kondo temperature on the Cu(111)
surface contrasts with the behavior previously reported for the
Co/Cu(100) system, where a considerable increase of the Kondo
temperature (from 70-90K in tunneling to 150K in point contact)
was observed \cite{neel07}. As will be shown below this difference
can be ascribed to the sensitivity of the Kondo effect to the
local atomic geometry.

\par\

To obtain a physical understanding of the structural sensitivity,
we have modelled at first the atomic relaxation in the single Co
atom junction under the influence of the tip proximity and then
considered its consequence on the electronic structure. Indeed,
reducing the tip-substrate separation can induce a local
perturbation in the atomic ordering at the junction which can
affect the coupling between the orbitals of the electrodes and of
the Co atom and consequently the electronic and the magnetic
properties of the system. To simulate the nanoscale-junction on
the atomic scale, we have performed molecule static (MS)
calculations with many-body interatomic potentials
\cite{levanov00}. In these simulations the tip has been
represented as a pyramid consisting of 10 Cu atoms arranged in
fcc(111) stacking. Figure ~\ref{fig2} shows the variation of the
tip-adatom and the adatom-substrate separations during the tip
displacement (panel a and b, respectively). On a first glance one
can see that beside an initial region, the tip- Co atom as well as
the Co atom-substrate distances are not linearly proportional to
the tip displacement. As the tip-substrate distance is reduced,
the atomic order at the junction relaxes: the atoms of the tip,
the Co impurity as well as the atoms of the substrate move to new
equilibrium positions. The real tip-substrate distance is then a
dynamic variable according to the specific location of the tip and
to its attractive and repulsive interaction with the surface and
the impurity. Specifically, up to the minimum distance of 5.3\AA,
the tip-Co atom distance is almost linear with the tip
displacement. Approaching further, the distance between the
opposite sides of the nanometer scale junction is reduced to a
larger extend than the effectively applied tip displacement due to
an attractive interaction (up to 4.7\AA). Reducing the
tip-substrate distance below 4.7\AA, the interaction becomes
repulsive. At this tip proximity, the adatom-substrate distance,
defined as the vertical distance between the Co adatom and its
first nearest neighbor, is strongly reduced while the distance
between the tip and adatom is only slightly decreased. This
implies that the Co adatom shifts towards the substrate. As a
consequence when the point contact configuration is reached, the
Co-Cu(111) surface distance compares to the equilibrium distance
predicted for the tunneling condition (dotted line in panel b).

Figure \ref{fig2} compares also the atomic relaxation process on
the two copper surfaces. The general trend of attractive and
repulsive interaction of tip-atom and surface can be observed in
both cases. However, differences in the atom dynamics under the
influence of the tip and in the Co-surface distance are obvious.
Specifically,  under the influence of the tip the Co impurity is
pushed deeper into the Cu(100) surface in point contact
configuration than it is in tunneling conditions (black dotted
line in panel b). Therefore, it can be expected that the stronger
interaction with the surface increases on the Cu(100) substrate
the hybridization of the d-levels of the Co adatom with the sp
states of the surface.

\begin{figure}[t]
 \centering
   \includegraphics[draft=false,keepaspectratio=true,clip,width=0.95\linewidth]
   {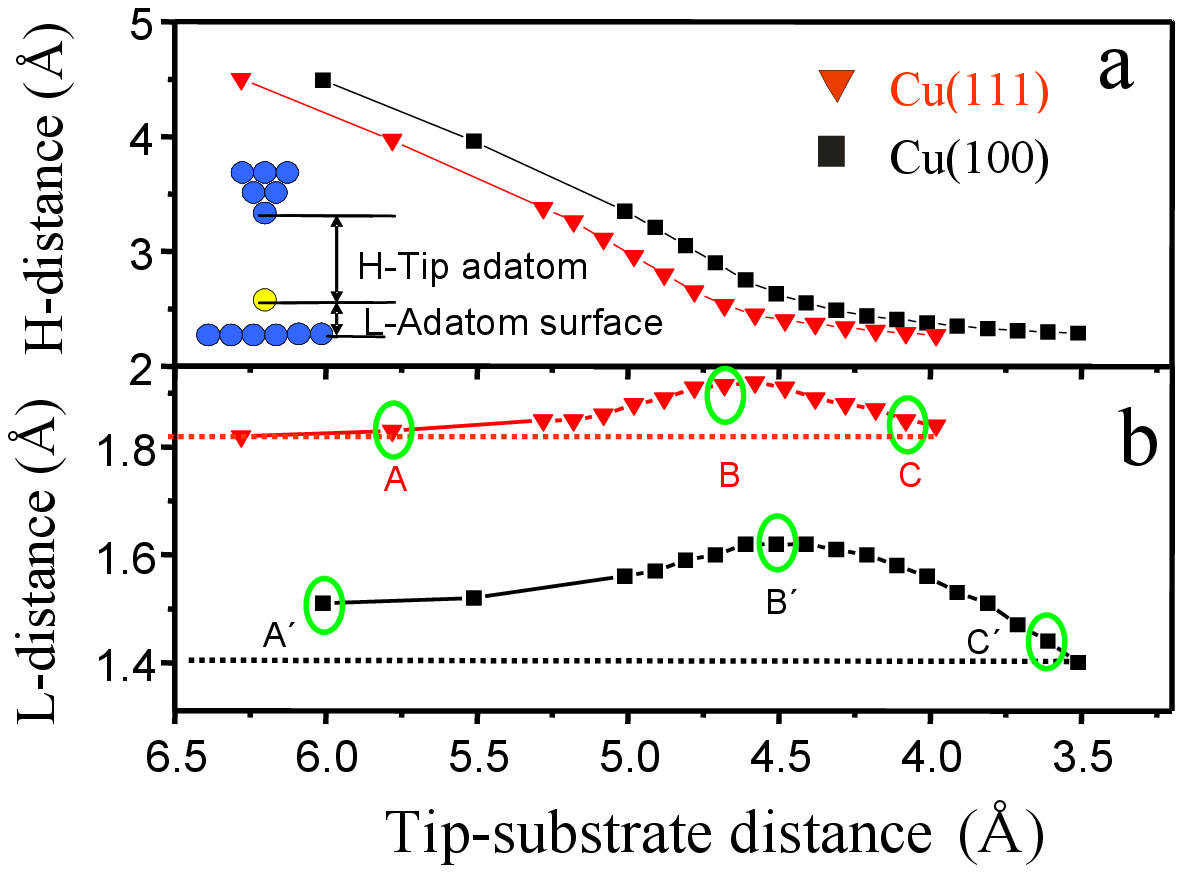}
   \caption{\label{fig2} Atomic relaxation at the single Co atom junction
   as a function of the tip-substrate displacement.
   The tip-adatom (H) and the adatom-substrate (L)
distances for Co adatom on Cu(111) and Cu(100) surface are shown
in panel a and b, respectively. The dotted lines emphasize the
relative Co-substrate distance in point contact and in tunneling
conditions. A, B and C indicate the position where the LDOS shown
in figure 3 have been calculated. }
 \label{fig2}
\end{figure}

This and its consequence on the magnetic properties of the
junction at various tip proximity, can be understood calculating
the local density of states (LDOS) of the Co adatom. All the
calculations were performed within the linear combination of
atomic orbital (LCAO) formalism by means of density functional
theory (DFT) implemented in SIESTA\cite{soler02}. The geometry was
optimized by SIESTA until all residual forces on each atom are
smaller than 0.01eV/\AA~\cite{MD-Sie,siesta}.

In figure ~\ref{fig3} the d-levels of Co/Cu(111) is shown for
different tip-substrate separations (denoted A, B and C in
figure~\ref{fig2}) with the energies given with respect to the
Fermi level. It can be seen that only the occupied density of
states of the Co adatom on Cu(111) are slightly affected by the
tip substrate distance. Moreover, the energy difference between
the center of the occupied spin-up band (or majority states) and
the center of the partially unoccupied spin-down band (or minority
states) U for three tip-substrate separations are nearly the same.
Consistently, also the magnetic moment of the Co adatom at these
three tip-substrate separations (1,99$\mu_{B}$, 1.96$\mu_{B}$ and
1,78$\mu_{B}$, respectively) are only slightly affected by the tip
proximity. On the contrary a large energy shift of the d-levels
was reported for Co adsorbed on Cu(100) surface \cite{neel07,
huang06}. A comparison of the energy position of the occupied
d-levels is shown in figure \ref{fig3}b. On both surfaces the
position of the occupied d-levels shifts towards higher energies
under the influence of the tip proximity. On Cu(111) surface this
shift is, however, much smaller. On Cu(100) the substantial change
in the occupation of the d-levels is reflected in the increase of
the Kondo temperature in point contact. Accurate calculations of
the expected increase of the Kondo temperature on the these
surfaces is, however, not straight forward. Nonetheless, the
theoretical predictions and the experimentally observed Kondo
temperature in point contact regime follows the trend described by
the model proposed by Wahl \emph{et al}. \cite{wahl04} for the
tunneling regime. The increase of the occupation of the d-level
effects sensibly the Kondo temperature on Cu(100) and almost
negligibly on the Cu(111) surface.

\begin{figure}[t]
 \centering
   \includegraphics[draft=false,keepaspectratio=true,clip,width=1\linewidth]{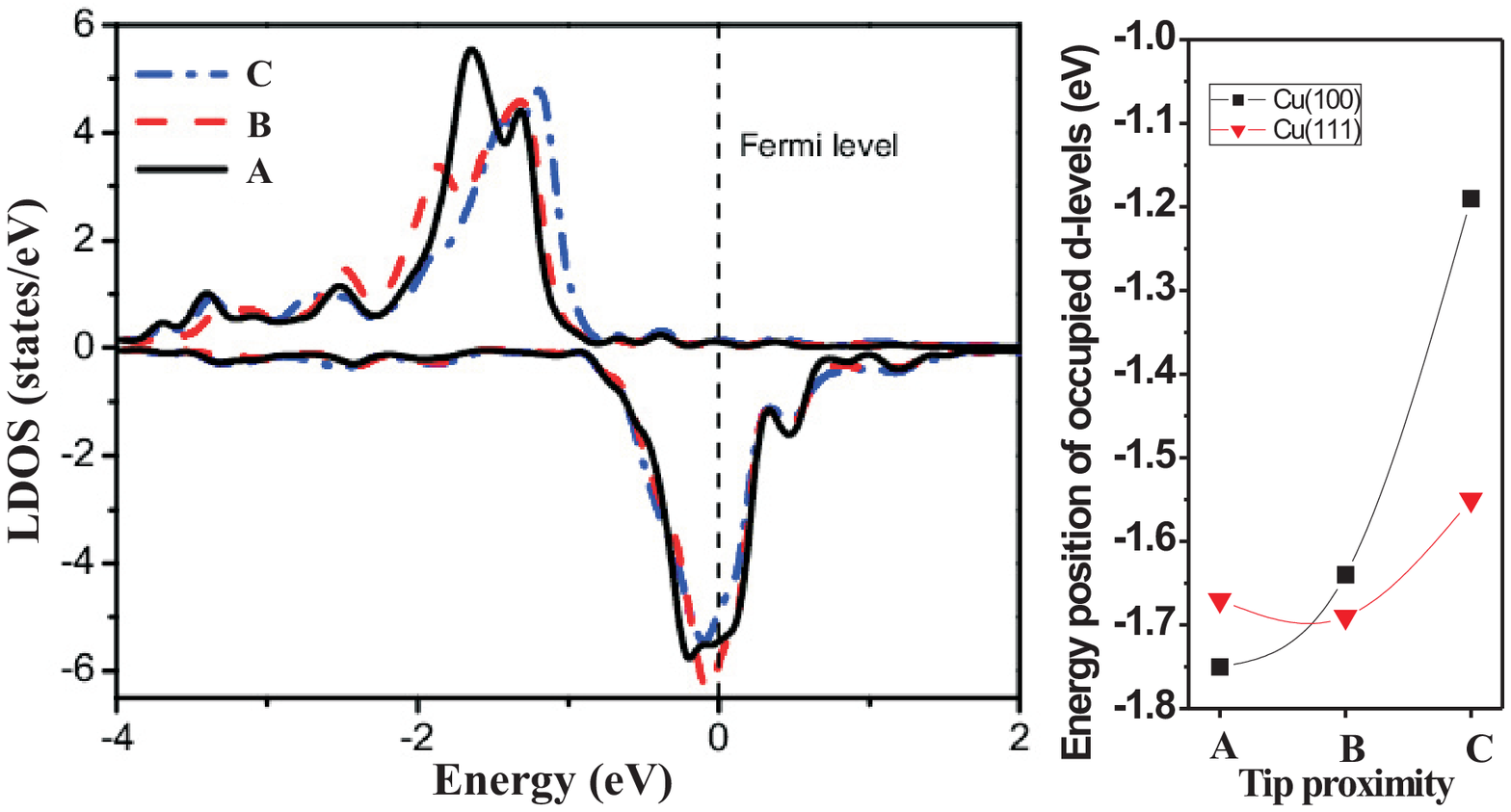}
   %bb=43 215 650 605
   \caption{\label{fig3}  Influence of the tip-proximity on the d-levels of Co atoms on
   Cu(111).
   a) Spin polarized LDOS for the d-levels.
   The curves are calculated for
   tip-substrate displacements as denoted in figure~\ref{fig2}.
   b) Energy position of the occupied d-levels on Cu(111) and Cu(100) at different tip-substrate positions.
   The lines connecting the points are a guide to the eye.}

 \label{fig3}
\end{figure}

\par\

In conclusion, the present experimental and theoretical work
demonstrates that the local atomic geometry plays a major role in
the electron transport properties of nanoscale junctions. The tip
proximity in the point contact regime influences the atomic
relaxation in the single atom junction and thereby determines the
lattice equilibrium position. These structural relaxations induce
a modification of the sp-d hybridization between the electrode
surface and the bridging atom. While on the closed packed Cu
surface the impurity d-level is less affected, it shifts
substantially on the open (100) surface. This explains the
striking difference observed in the behavior of the Kondo
temperature of Co adatoms upon point contact formation on Cu(111)
and Cu(100). We believe that these results have general validity
and might clarify a few of the uncertainties in the electron
transport through nanometer scale junctions characterized by
break-junction experiments.

We acknowledge P.Wahl for fruitful discussions. The work was
supported by Deutsche Forschungsgemeinschaft (DPG SPP 1165, SPP
1243 and SSP 1153)


\begin{thebibliography}{}
\bibitem{siesta}
LDA approaches have been used for the exchange and correlation
potential, and a 250 Ry energy cutoff is used to define real-space
grid for numerical calculations involving the electron density.
Core electrons of all elements are replaced by nonlocal
norm-conserving Troullier-Martins pseudopotentials. Valence
electrons of the Cu substrate are described using double-$\zeta$
plus polarization atomic orbital basis set, and a triple-$\zeta$
plus polarization atomic orbital basis sets for Co adatom.
\bibitem{MD-Sie}
We chose three different tip-substrate distances (denoted A, B and
C in figure~\ref{fig2}) and perform a fully relaxation by Siesta
code. The displacements of the Co adatom at three tip-substrate
separations (denoted $a$, $b$ and $c$ in Fig.~\ref{fig2})
calculated by the MS method are 0.02\AA, 0.21\AA~ and 0.05\AA,
respectively. While these are 0.01\AA, 0.26\AA and 0.08\AA
~obtained from Siesta code, coinciding with the MD results very
well. Relaxation of the atoms in the tip is also taken into
account in our calculations~\cite{huang06}.
\end{thebibliography}
\end{document}